\documentclass[aps,prl,twocolumn]{revtex4-1}

\pdfoutput=1
\usepackage[breaklinks]{hyperref}
\usepackage{graphicx}
\usepackage[applemac]{inputenc}
\usepackage{natbib}
\usepackage{bm}
\usepackage{amssymb}
\usepackage{amsbsy}
\usepackage[usenames]{color}
\usepackage{mathrsfs,mathcomp}
 \usepackage[normalem]{ulem}
\usepackage{amsmath}
\usepackage{amsfonts}
\usepackage{mathrsfs}
\usepackage{subfigure}
\usepackage{ifsym} 
\usepackage{multirow}
\usepackage{textcomp}
\graphicspath{{./figuresPDF/}}

\begin{document}

\title{Universality of Tip Singularity Formation in Freezing Water Drops}

\author{A.G. Mar\'in$^1$, O.R. Enr\'iquez$^2$, P. Brunet$^3$, P. Colinet$^4$, and J.H. Snoeijer$^{2,5}$} 

\affiliation{$^1$Institut f\"ur Str\"omungsmechanik und Aerodynamik, Bundeswehr University Munich, Germany\\
$^2$Physics of Fluids Group, Faculty of Science and Technology, Mesa+ Institute, University of
Twente, 7500 AE Enschede, The Netherlands \\
$^3$Laboratoire Mati\`ere et Syst\`emes Complexes UMR CNRS 7057, 10 rue Alice Domon et L\'eonie Duquet 75205 Paris Cedex 13, France\\
$^4$Universit\'e Libre de Bruxelles, Laboratory TIPs (Transfers, Interfaces and Processes) Fluid Physics Unit, CP 165/67, 
Av. F.D. Roosevelt, 50, 1050 Brussels, Belgium\\
$^5$Department of Applied Physics, Eindhoven University of Technology, P.O. Box 513,
5600 MB Eindhoven, The Netherlands}

\date{\today}

\begin{abstract}
A drop of water deposited on a cold plate freezes into an ice drop with a pointy tip. While this phenomenon clearly finds its origin in the expansion of water upon freezing, a quantitative description of the tip singularity has remained elusive. Here we demonstrate how the geometry of the freezing front, determined by heat transfer considerations, is crucial for the tip formation. We perform systematic measurements of the angles of the conical tip, and reveal the dynamics of the solidification front in a Hele-Shaw geometry. It is found that the cone angle is independent of substrate temperature and wetting angle, suggesting a universal, self-similar mechanism that does not depend on the rate of solidification. We propose a model for the freezing front and derive resulting tip angles analytically, in good agreement with observations.

%

\end{abstract}

\maketitle


%

\begin{figure}[h!]
\includegraphics[]{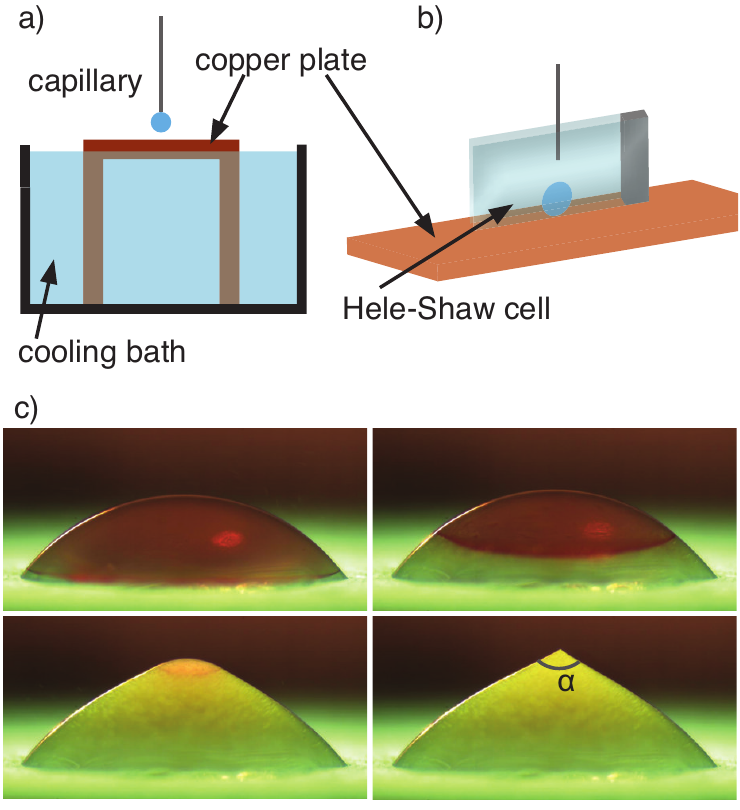}
\caption{(color online) Experimental set-up. (a) Water drops are deposited on a copper plate immersed in a cooling bath of ethylene-glycol, ethanol and dry ice. (b) The shape of the advancing freezing front can be observed using a 2D-like Hele-Shaw set-up. (c) A freezing water droplet with red dye. The position of the tri-junction point is clearly visible, while the images give also give a qualitative impression of the geometry of the freezing front.}\label{fig1}
\end{figure}

Liquid solidification can lead to intricate morphological structures, from dendritic growth \cite{Glicksman2011} to the fascinating complexity of snowflakes \cite{snowflake}. In an apparently much simpler situation of a water drop freezing on a cold substrate, it has been observed that the final shape of the ice drop is pointy \cite{Anderson_JCG96,Ajaev2003,Snoeijer:2012ft,Enriquez:2012hu}, with a sharp tip that is reminiscent of the domes of orthodox churches (Fig.~\ref{fig1}). Intriguingly, the sharp tip appears despite the presence of liquid surface tension, which usually tends to smooth out sharp features. These singular ice drops can be observed in frozen water accretion on aircraft cabins during flights \footnote{See, for example, SKYbrary, ``In-flight icing,'' \url{<http://www.skybrary.aero/index.php/In-Flight_Icing>}}, during solidification for freeze drying purposes \cite{mellor1978fundamentals}, and in recent studies on supercooling \cite{Jung:2012hs} and icing of substrates \cite{Jung:2011ky}. A similar mechanism is thought to be at the origin of the formation of spiky micro-structures following the irradiation by high-power ultra-short lasers in Germanium and Silicium substrates \cite{Kolasinski_COSSMS07}, which like water are materials that expand upon freezing.

\begin{figure*}[t]
\includegraphics[width=2 \columnwidth]{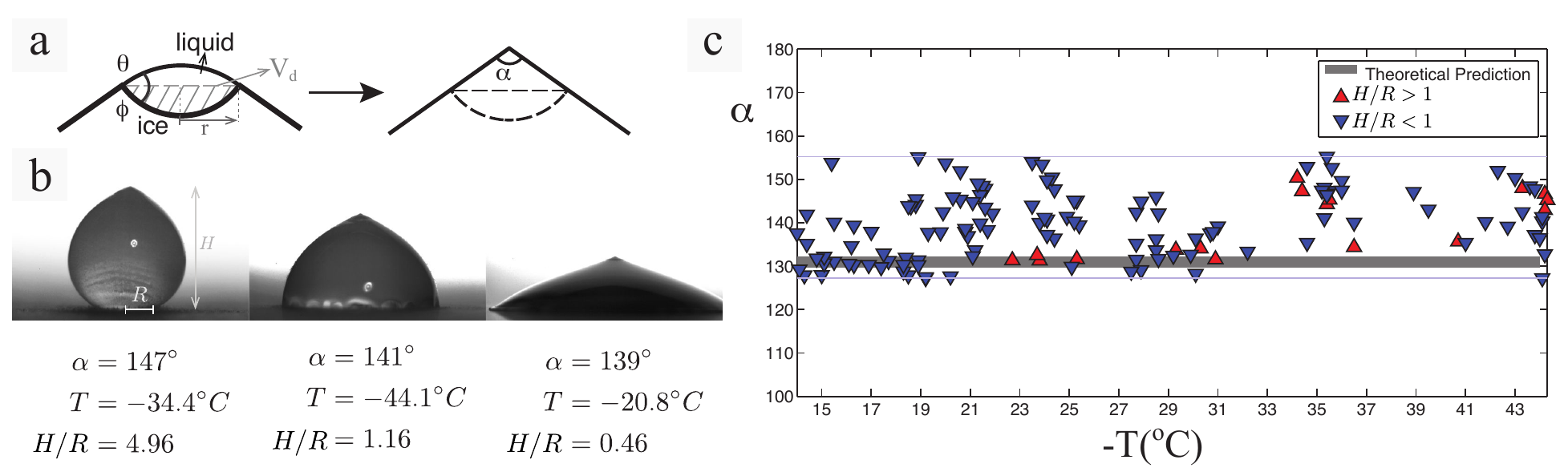}
\caption{(color online) Tip formation on freezing drops. (a) Sketch of the geometry during the final stages of the freezing process. We define the angles and angles $\alpha,\theta,\phi$, the radius $R$, and the ``downward volume" $V_d$. (b) Different contact angles (i.e. aspect ratios) achieved at different temperatures. (c) Measurement of the tip angle $\alpha$ for the full range of temperatures explored. Base-down triangles represent data from droplets at high contact angles and therefore a height-to-radius aspect ratio $H/R>1$. Base-up triangles represent drops with low contact angle characterized by aspect ratios $H/R<1$. We find no systematic dependence of $\alpha$ with global drop shape and substrate temperature. The grey line indicates the theoretical prediction (\ref{eq:alpha}).}\label{fig2}
\end{figure*}

Though the formation of pointy ice drops has been attributed to the expansion of water upon freezing, there is still no satisfactory explanation for this phenomenon. Previous studies revealed that the freezing can indeed yield a tip singularity, by modeling a planar solidification front reaching the top of the drop \cite{Anderson_JCG96,Snoeijer:2012ft}. However, these theories predict a singularity only when the ratio of solid and liquid densities is below 0.75: this clearly does not explain the appearance of conical ice drops, since for water $\nu \equiv \rho_s /\rho_\ell=0.92$. The paradox can be resolved by assuming that the freezing dynamics induces a contact angle, with a slope discontinuity at the solid/liquid/air tri-junction point that depends on the freezing rate \cite{Anderson_JCG96}.
Alternatively, the singularity was also recovered for realistic $\nu$ when numerically treating the solidification dynamics in full detail, but without the assumption of a dynamic contact angle \cite{Ajaev2003, Ajaev2004}. 
The true mechanism behind the tip singularity has therefore remained elusive, in particular since there is a lack of systematic experiments to which any of the theories can be compared. 


In this Letter we reveal that the geometry of the freezing front, essentially determined by the final stages of a quasi-steady heat transfer problem, is responsible for the formation of pointy ice drops. First, we experimentally show that the cone angles at the tip are universal, and do not depend on the substrate temperature, excluding the influence of the solidification rate. Next, we reveal the boundary conditions of the solidification front by tracking the freezing process in a Hele-Shaw geometry. It is found experimentally, and explained theoretically from heat conduction, that the front develops a spherical shape that ends perpendicularly to the solid-air interface. Taking this into account into the mass balance during solidification, we then show how the singularity emerges for any density ratio $\nu < 1$. The theory predicts a cone angle $\alpha = 131^\circ$ for water drops [$\alpha$ defined in Fig.~\ref{fig1}(c)], which falls within the range of experimental observations.

\paragraph{Experiments.---}
Droplets of pure water (milli-Q, degassed) were frozen on a copper structure that is partially immersed in a cooling bath composed of ethylene-glycol, ethanol and dry ice [Fig. \ref{fig1}(a)]. With this mixture, the temperature can be controlled in the range -78 to -17\textdegree C by modifying the volume fraction of the two liquids \cite{Jensen2000}. Droplets of volume 4-8 \textmu l were deposited using a syringe pump (Harvard Apparatus PHD ULTRA) and a 200 \textmu m capillary. The temperature on the plate was measured near the droplet using a thermocouple. We focus here on temperatures above -44\textdegree C, for which reproducible experiments could be performed. At lower temperatures non-directional freezing and multiple freezing fronts appear. The freezing process was recorded, using a PCO camera (Sensicam QE), a long distance microscope (Edmund Optics VZM1000) and diffused back lighting, until the tip was formed.

We extracted the final shape of the drops through image analysis and fitted third-order polynomials to the left-hand and right-hand regions close to the tip. The angle of the tip is then computed as the intersecting angle of the polynomials, with an experimental error bar of $\pm 5^\circ$ on average. The total time for the solidification is on the order of 1 second for the coldest cases and 10 seconds for the warmest. Importantly, the liquid near the contact line freezes long before reaching the equilibrium contact angle, so that different contact angles can be achieved by varying the height of deposition. This results in drops of different aspect ratio, $H/R$, where $H$ is defined as the final height of the ice drop and $R$ the radius of the wetted area (Fig.~\ref{fig2}b). Various movies of the freezing process can be found as Supplementary material \cite{supplementary}.

Figure~\ref{fig2}b shows typical shapes of ice drops, as obtained for different temperatures and contact angles.  Despite the large disparity of drop shapes, the formation of the tip singularity appears to be independent of contact angle and substrate temperature. This can be inferred from Fig.~\ref{fig2}c, where we report the cone angle $\alpha$ for  more than 200 experiments, carried out at different temperatures (horizontal axis) and for different aspect ratios (upward/downward symbols). All measurements fall within a well-defined range of tip angles, characterized by an average and standard deviation $\alpha = 139^\circ \pm 8^\circ$. The data give no evidence for any correlation of $\alpha$ with temperature and aspect ratio. The experiments thus show that the tip formation is not influenced by global geometry of the drop, nor by the rate at which the solidification occurs -- the latter suggesting that the singularity is the outcome of a quasi-static process. 
The observed variability in $\alpha$ is beyond the accuracy of the measurement, and appears to be due to the conditions in which the experiments where performed, i. e. with the droplet exposed to air currents and vapor from the cooling bath.

\paragraph{Freezing front.---}

To obtain further insight in the tip formation, we next investigate the shape of the solidification front. The still images in Fig.~\ref{fig1}(c) suggest that the front does not remain planar, as was also discussed in~\cite{comment,reply_to_comment}. However, in order to achieve quantitative access to the advancing front, we constructed a Hele-Shaw cell with two microscope slides separated by a 1 mm spacer [Fig. \ref{fig1}(b)]. The cell was placed on the copper plate and the capillary carefully maneuvered between the walls. The gap is wide enough for the drop to form a conical tip, but this is not always visible due to the presence of a wetting meniscus. To minimize image distortion due to this meniscus, the glass is treated such that the wetting contact angle $\approx 90^\circ$. This gives a clear view on the quasi-two-dimensional freezing process -- typical videos can be found in the supplementary materials \cite{supplementary}. 

\begin{figure}[t]
\includegraphics[width=1 \columnwidth]{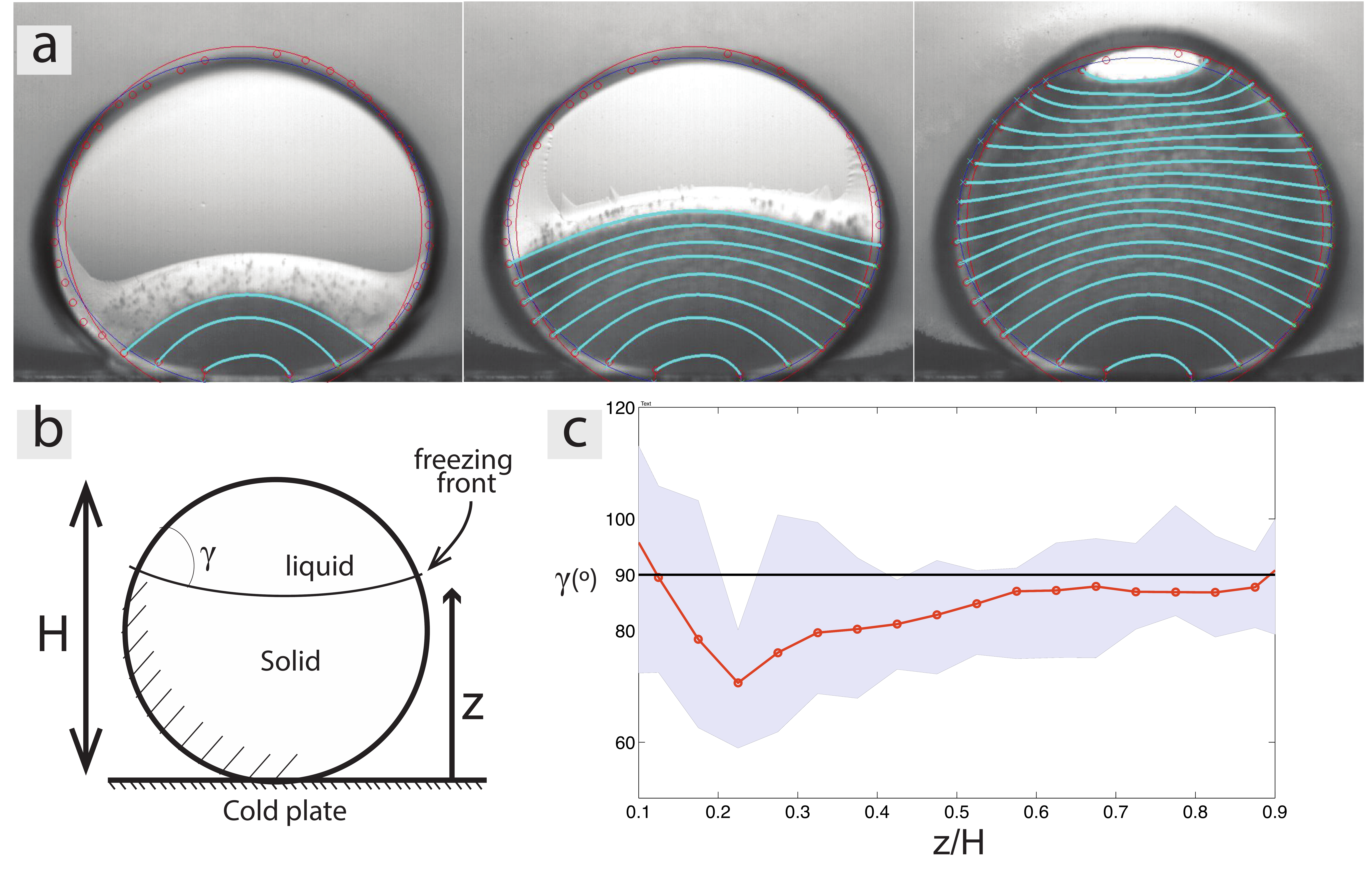}
\caption{(color online) Freezing experiments with droplets sandwiched in a Hele-Shaw cell. (a) Evolution of the freezing front (green line) in different stages of the process. (b) Sketch and definitions of angles and distances. (c) Front-to-interface angle $\gamma$ plotted against the relative height of the freezing front $z/H$. Within experimental variations, we find $\gamma \approx 90^\circ$.}\label{fig3}
\end{figure}

Figure \ref{fig3}(a) shows that the solidification front in the Hele-Shaw cell grows towards the top of the drop in a similar fashion as the unconfined experiment, although the timescale of the process is a bit faster. In the first instants of the process, vapor condensates on the glass slides resulting in a ``frost halo" \cite{Jung:2012hs} around the drop. Such event occurs simultaneously as the partial and kinetically-controlled recalescent freezing \cite{KavehpourAPS2012}, also visible in Figure \ref{fig3}(a) as a brighter area above the freezing front.

Our prime interest here is to extract quantitative information on the geometry of the freezing front. The front shown in Fig.~\ref{fig3}(a) has a convex shape at the early stage of the freezing, while at the last stages the curvature is inverted towards a concave geometry. Interestingly, these profiles closely resemble two-dimensional numerical simulations \cite{Ajaev2003}. At all times, the experimentally observed freezing front appears to be perpendicular to the ice-air interface. This is confirmed in Fig.~\ref{fig3}(c), where we present the angle $\gamma$ defined in Fig.~\ref{fig3}(b), as a function of the height $z$ of the tri-junction point. The red line corresponds to the average over 20 experiments, performed at temperatures ranging from -30 to -15\textdegree C. We find that the front is nearly perpendicular during the entire experiment. During the final stages we find an average and standard deviation $\gamma = 87^\circ \pm 8^\circ$. 

\paragraph{Heat-transfer-limited self-similar freezing dynamics.---}
Coming back to the axisymmetric case we now derive the shape of the solid-liquid front from the heat transfer in the late stages of the freezing process [cf. Fig. \ref{fig2}(a)]. We neglect any small-scale kinetic undercooling or Gibbs-Thomson effect (as considered e.g. in \cite{Ajaev2004}), such that the front here always remains at the equilibrium melting temperature $T_m$. As the air surrounding the drop has a much smaller thermal conductivity than the solid (and the liquid), the latent heat released by the advancing front must be evacuated via the solid, while the liquid remains at uniform temperature $T_m$. The fact that heat cannot cross the solid-air boundary has an important consequence: it implies that the isotherms, in particular the the freezing front at $T_m$, are locally perpendicular to the solid-air boundary, i.e. $\gamma = 90^\circ$. This simple argument is in good agreement with the Hele-Shaw cell experiments, even though heat transfer via the glass slides might not be entirely negligible there.  

The final stages of the heat transfer are expected to be self-similar. Namely, the angle $\gamma$ made by the front with the external surface remains approximately constant, and the freezing is characterized by a single length scale, $r$, the radius at the triple junction. Based on this length, one can derive a scaling-law for the normal velocity $v_n$ of the front, which is proportional to the rate at which latent heat released by the front is evacuated. This gives $v_n \sim -dr/dt \sim \lambda_s \delta T/\rho_s {\cal L}_m r$, where $\lambda_s$ is the solid thermal conductivity, ${\cal L}_m$ is the latent heat of melting, and the undercooling $\delta T=T_m-T$. Here we assumed that the heat transfer process in the solid is quasi-steady, i.e. that the time scale of front motion $r/v_n$ is much larger than the thermal diffusion time scale $r^2/\kappa_s$, where $\kappa_s$ is the thermal diffusivity of the solid. Taking into account that $\kappa_s=\lambda_s/\rho_s c_{p,s}$, where $c_{p,s}$ is the thermal capacity of the solid, this is equivalent to assuming that the Stefan number $S=c_{p,s} \delta T/{\cal L}_m$ is small \cite{Ajaev2003}. Actually, we find $S=0.27$ for the maximal value of $\delta T=44$ K in our experiments, such that this quasi-steadiness assumption is sufficiently accurate here [as also confirmed in Fig.~\ref{fig2}(c)]. Note that solving the above energy balance for $r$ leads to a classical $r^2$-law, i.e. $r^2$ decreases linearly with time during tip formation.

The above theory also provides the self-similar shape of the freezing front, which is crucial for understanding the tip formation. The heat transfer problem amounts to solving $\nabla^2 T=0$ in the solid, with the boundary condition that the isotherms make an angle $\gamma = 90^\circ$. When the solid approaches a conical shape, the resulting isotherms are portions of concentric spheres centered at the final cone tip [Fig. \ref{fig2}(a)]. The two-dimensional equivalent is that the freezing front is a portion of a circle -- in good agreement with the concave shape in the final stage [Fig. \ref{fig3}(c)]. This solution is stable with respect to dendrite formation given that cooling is from the solid side \cite{Glicksman2011}. Hence, the front will remain spherical during the self-similar final stages of tip formation.



\paragraph{Geometric theory for tip formation.---}

The results above point towards a scenario where the tip is formed by a quasi-static mechanism. Based on this, we propose a model in the spirit of  \cite{Anderson_JCG96}, but taking into account the self-similar spherical geometry of the freezing front. The starting point is mass conservation 

\begin{equation}\label{eq:mass}
\frac{d}{dz}\left( V_\ell + \nu V_s \right) = 0, 
\end{equation}
here expressed in terms of liquid and solid volumes $V_{\ell,s}$ and density ratio $\nu$. Since temporal dynamics is unimportant, the conservation law has been written in terms of a derivative with respect to $z$, the height of the tri-junction. The total liquid volume can be decomposed into a spherical cap of angle $\theta$ and a downward volume $V_d$ [cf. Fig.~\ref{fig2}(a)]. The liquid and solid volumes then are

\begin{equation}\label{eq:volume}
V_\ell = r^3 f(\theta) + V_d; \quad V_s = - V_d + \int_0^z dz'\, \pi r(z')^2,
\end{equation}
where $r(z)$ and $\theta(z)$ are the local radius and angle of the frozen drop, and the geometry of a spherical cap gives

\begin{equation}\label{eq:cap}
f(\theta) = \frac{\pi}{3}\left( \frac{2-3\cos \theta +\cos^3 \theta}{\sin^3\theta}\right).
\end{equation}
Closing the problem requires an expression for $V_d$, which in general implies a full solution of the solidification front. For the final stages, however, we can take advantage of our previous observation that the front develops a spherical shape with downward angle $\phi=\gamma-\theta$, such that $V_d = r^3 f(\gamma - \theta)$. With this, (\ref{eq:mass}--\ref{eq:cap}) give a closed set of equations for $r(z)$, indeed predicting a sharp tip as $r \rightarrow 0$ \footnote{The $z$-derivative of (\ref{eq:mass}) gives terms $d\theta/dz$ and $dr/dz$, the latter being $-1/\tan \theta$. The tip formation is obtained by taking $r \rightarrow 0$: this gives $d\theta/dz=0$, which is achieved for the condition given as (\ref{eq:alpha})}. At the singularity, $\theta$ obeys

\begin{equation}\label{eq:alpha}
f(\gamma - \theta) + f(\theta) = \nu \left[ f(\gamma-\theta) + \frac{\pi}{3} \tan \theta \right],
\end{equation}
from which we can infer $\alpha= \pi - 2\theta$, for any density ratio. 

The central result of this analysis is that, combined with our result that $\gamma=90^\circ$, equation (\ref{eq:alpha}) gives a parameter-free prediction for the tip angle of ice drops: $\alpha = 131^\circ$. This is consistent with our experimental observation $\alpha = 139^\circ \pm 8^\circ$, though most experiments are slightly above the theoretical prediction. We tentatively attribute the experimentally observed variability in the cone angle to variations of $\gamma$. Inserting the $\gamma$-variations measured in the Hele-Shaw cell in (\ref{eq:alpha}) indeed gives $\alpha = 133^\circ \pm 5^\circ$, in close agreement with experiments. 

Equation (\ref{eq:alpha}) has an elegant interpretation in terms of the volumes before and after freezing. After multiplication by $r^3$, the left hand side represents the unfrozen liquid volume, consisting of two spherical caps. This mass is to be transformed into ice, where due to expansion factor $\nu$ the upward liquid sphere is transformed into a cone of volume $\frac{\pi}{3}r^3  \tan \theta$. The tip angle is thus determined from purely geometrical considerations. Interestingly, the model does not display a critical density ratio: a tip is formed even for $0< 1-\nu \ll 1$. This can be seen by expansion of (\ref{eq:alpha}), yielding a nonzero angle $\theta = \frac{12}{\pi}f(\gamma)(1-\nu)$, in radians. This is in marked contrast with the model where the solidification front was assumed planar \cite{Anderson_JCG96,Snoeijer:2012ft}, which predicted tip formation only for $\nu$ below a critical value $3/4$.

\paragraph{Outlook.---}
To summarize, we have experimentally and theoretically identified the geometry of the freezing front and demonstrated its crucial role for the formation of conical ice drops. At the macroscopic scales considered here, the final cone angle is found to be independent of the history and rate of freezing. For future work it would be particularly interesting to further explore the physics of the singularity: down to what length scale does the tip of the drop remain sharp? Due to the vanishing length scale and diverging rate of solidification, $dr/dt \sim 1/r$, kinetic undercooling and the Gibbs-Thomson effect could come into play \cite{Ajaev2004}. We expect that these (or other) microscopic effects will ultimately regularize the singularity. This would give insight in solidification under extreme and more realistic conditions \cite{Jung:2011gf}. More generally, the present work already reveals that understanding the boundary conditions at the triple line are of decisive importance.

\paragraph{Acknowledgments.---} AGM and ORE contributed equally on the paper. We thank Koen Winkels for many discussions. JHS acknowledges financial support from NWO through VIDI Grant No. 11304. PC acknowledges financial support from FRS-FNRS and BELSPO (IAP 7/38 $\mu$-MAST).



\bibliographystyle{unsrt} 
\bibliography{bibfreeze}

\end{document}